# Correction of Blurring due to a Difference in Scanning Direction of Field-Free Line in Projection-Based Magnetic Particle Imaging

Kenya Murase*, Kazuki Shimada, and Natsuo Banura

Department of Medical Physics and Engineering, Division of Medical Technology and Science, Faculty of Health Science, Graduate School of Medicine, Osaka University, Suita, Osaka 565-0871, Japan

* Corresponding author, email: murase@sahs.med.osaka-u.ac.jp

**Abstract**

In projection-based magnetic particle imaging (MPI) with a field-free-line (FFL) encoding scheme, projection data are usually acquired by moving the FFL in a zigzag and a difference in the projection data occurs depending on the scanning direction of FFL, resulting in blurring in the reconstructed images. In this study, we developed a method for correcting the blur by deconvolution using a signal-delay constant ($\xi$). The $\xi$ value for correction ($\xi_c$) was determined by acquiring projection data in positive and negative directions and searching for the $\xi$ value which minimized the 2-norm between the deconvolved projection data in the two directions. We validated our method using a line and A-shaped phantoms for various velocities of FFL ($v_{FFL}$). The $\xi_c$ value correlated linearly with $v_{FFL}$. The full width at half maximum of the line phantom decreased significantly after correction of the blur. The effectiveness of our method was also confirmed by the MPI images of the A-shaped phantom. These results suggest that our method will be useful for enhancing the reliability of projection-based MPI.

## I. Introduction

Magnetic particle imaging (MPI) is a recently introduced imaging method [1] that allows imaging of the spatial distribution of magnetic nanoparticles (MNPs) with high sensitivity, high spatial resolution, and high imaging speed. MPI uses the nonlinear response of MNPs to detect their presence in an alternating magnetic field (drive magnetic field). Spatial encoding is accomplished by saturating the MNPs over most of the imaged region using a static magnetic field (selection magnetic field), except in the vicinity of a field-free point [1] or field-free line (FFL) [2].

More recently, we developed a system for projection-based MPI with an FFL encoding scheme [3]. In such a system, projection data are usually acquired by moving the FFL in a zigzag in order to make the acquisition time as short as possible, which can cause a difference in the projection data depending on the scanning direction of the FFL [4], resulting in blurring in the reconstructed images. When considering the practical application and enhancement of the reliability of projection-based MPI, it is necessary to correct for such a blur. The purpose of this study was to develop a method for its correction, and to investigate the validity and usefulness of this method by phantom experiments.

## 2. Materials and Methods

### 2.I. MPI System

In this study, we used the Osaka MPI scanner II, which is an extended version of our previous scanner [3]. In brief, a drive magnetic field was generated using an excitation coil, which was controlled with a sinusoidal wave generated using a digital function generator. The frequency and peak-to-peak strength of the drive magnetic field were 400 Hz and 20 mT, respectively. The signal generated by MNPs was received by a gradiometer coil, and the third-harmonic signal was extracted using a lock-in amplifier. The output of the lock-in amplifier was converted to digital data by a personal computer connected to a multifunction data acquisition device. The selection magnetic field was generated by two opposing neodymium magnets (500 mm in height, 122 mm in width, and 67 mm in thickness). The FFL can be generated at the center of the two neodymium magnets. The gradient strength of the FFL is 3.9 T/m, 0.1 T/m, and 3.7 T/m, in the $x$, $y$, and $z$ axes [3], respectively.

To acquire projection data for image reconstruction, a sample located in the receiving coil was automatically rotated around the z axis and translated in the x axis using an XYZ-axes rotary stage controlled using LabVIEW (National Instruments Co., TX, USA). In this study, projection data were acquired by rotating both the sample and receiving coil simultaneously over 180° in steps of 5°. For each projection angle, projection data were acquired by translating the sample and receiving coil simultaneously from −16 to 16 mm in the horizontal direction ($x$ axis) at 1 mm intervals, and each set of projection data was then transformed into 64 bins by linear interpolation. Transverse images were reconstructed from the projection data using the ML-EM algorithm over 30 iterations [3].

## 2.2. Correction of Signal Delay

The signal delay in projection data was modelled as follows. First, we assumed that the MPI signal at position $x$ and time t ( $S(x,t)$ ) is given by [4]

$$\frac{\partial S(x,t)}{\partial t} = -\frac{S(x,t) - S_{adiab}(x,t)}{\tau} \quad (1)$$

where $S_{\text{adiab}}(x,t)$ represents the adiabatic signal, i.e., the signal without delay and $\tau$ is a delay time constant. Transforming (1) yields

$$\frac{\partial S(x,t)}{\partial t} = \frac{dS(x)}{dx}\frac{dx}{dt} = \frac{dS(x)}{dx} v_{FFL} \quad (2)$$

where $v_{\text{FFL}}$ represents the velocity of the FFL. Thus, we get the following equation from (1) and (2):

$$\frac{dS(x)}{dx} = -\frac{S(x) - S_{adiab}(x)}{\xi} \quad (3)$$

where $\xi = \tau v_{FFL}$. We call $\xi$ the "signal-delay constant". This parameter has a unit of mm. Solving (3) yields

$$S(x) = \frac{1}{\xi}\int_0^x S_{\text{adiab}}(x') e^{-\frac{x-x'}{\xi}} dx' = S_{\text{adiab}}(x) \otimes \frac{1}{\xi} e^{-\frac{x}{\xi}} \quad (4)$$

where $\otimes$ denotes the convolution integral. When expressing (4) in a discrete and matrix form, (4) is reduced to the following equation:

$$\mathbf{S} = \mathbf{A}\mathbf{S}_{adiab} \quad (5)$$

where

$$\mathbf{S} = \left[S(x_1) \quad S(x_2) \quad \cdots \quad S(x_n)\right]^T \quad (6)$$

$$\mathbf{S}_{\text{adiab}} = \left[S_{\text{adiab}}(x_1) \quad S_{\text{adiab}}(x_2) \quad \cdots \quad S_{\text{adiab}}(x_n)\right]^T \quad (7)$$

and

$$\mathbf{A} = \frac{\Delta x}{\xi}\begin{pmatrix} e^{-\frac{x_1}{\xi}} & 0 & \cdots & 0 \\ e^{-\frac{x_2}{\xi}} & e^{-\frac{x_1}{\xi}} & \cdots & 0 \\ \cdots & \cdots & \cdots & \cdots \\ e^{-\frac{x_n}{\xi}} & e^{-\frac{x_{n-1}}{\xi}} & \cdots & e^{-\frac{x_1}{\xi}} \end{pmatrix} \quad (8)$$

In (6) and (7), $T$ denotes the transpose of a matrix. $\Delta x$ in (8) denotes a sampling interval.

With singular value decomposition (SVD), the matrix $\mathbf{A}$ in (5) can be expressed as the product of an $n\times n$ column-orthogonal matrix $\mathbf{U}$, an $n\times n$ diagonal matrix $\mathbf{W}$ and the transpose of an $n\times n$ orthogonal matrix $\mathbf{V}$ [5], i.e.,

$$\mathbf{A} = \mathbf{U}\mathbf{W}\mathbf{V}^T = \mathbf{U}[\text{diag}(w_i)]\mathbf{V}^T \quad (9)$$

where $w_i$ ($i$=1,2,⋯,$n$) are the diagonal elements of $\mathbf{W}$ (the singular values) which are nonnegative and can be ordered such that $w_1 \geq w_2 \geq \cdots \geq w_n \geq 0$. Thus, the matrix $\mathbf{S}_{\text{adiab}}$ can be calculated as

$$\mathbf{S}_{\text{adiab}} = \mathbf{A}^{-1}\mathbf{S} = \mathbf{V}[\text{diag}(1/w_i)](\mathbf{U}^T\mathbf{S}) \quad (10)$$

If $w_i$ was smaller than the maximal value of $w_i$ multiplied by a threshold value, 1/ $w_i$ in (10) was replaced by zero. In this study, the threshold value was fixed at 0.1.

To obtain the ξ value for correction ($\xi_c$), we acquired projection data in positive and negative directions in the horizontal ($x$) axis. We then performed the following calculation:

$$\xi_c = \arg\min_{\xi} \left\| \mathbf{S}_{\text{adiab}}^{\text{P}} - \mathbf{S}_{\text{adiab}}^{\text{N}} \right\|_2 \quad (11)$$

where $\mathbf{S}_{\text{adiab}}^{\text{P}}$ and $\mathbf{S}_{\text{adiab}}^{\text{N}}$ denote the deconvolved MPI signals in positive and negative directions, respectively. Actually, we calculated the 2-norm ($\left\| \mathbf{S}_{\text{adiab}}^{\text{P}} - \mathbf{S}_{\text{adiab}}^{\text{N}} \right\|_2$) for the $\xi$ value ranging from 0 to 2 mm in steps of 0.01 mm and then determined the $\xi$ value minimizing the 2-norm as $\xi_c$.

## 2.3. Phantom Experiments

To validate our method, we performed experiments using two kinds of phantoms (line and A-shaped phantoms) in which Resovist® was used as a source of MNPs.

The line phantom comprised a silicon tube (2 mm in diameter and 10 mm in length) filled with Resovist® at an iron (Fe) concentration of 500 mM. To investigate the effect of the speed of FFL, we varied $v_{\text{FFL}}$ as 0.62, 1.25, 1.67, and 2.0 mm/s and analyzed the correlation between $\xi_c$ and $v_{\text{FFL}}$ using linear regression analysis. It should be noted that in our MPI scanner, the position of the FFL is fixed, whereas the phantom is translated and rotated. Thus, the translational velocity of the phantom corresponds to $v_{\text{FFL}}$. The horizontal and vertical profiles along the lines passing through the center of the MPI image were calculated and the full width at half maximum (FWHM) was calculated by fitting the profiles by Gaussian function. The statistical significance in FWHM between before and after correction of the blur was analyzed by the paired Student's $t$-test, and a $P$ value less than 0.05 was considered statistically significant.

We also validated our method using an A-shaped phantom. This phantom consisted of silicon tubes 1.5 mm in diameter and filled with 500 mM Fe Resovist®.

## 3. Results

Figure 1(a) shows an example of the projection data acquired in positive (solid line) and negative directions (dotted line) for $v_{FFL}$ = 2.0 mm/s, demonstrating that there exists some shift between the two sets of projection data. Figure 1(b) shows an example of the 2-norm values calculated from (11) as a function of $\xi$ for $v_{FFL}$ = 2.0 mm/s. In this case, the $\xi_c$ value was calculated to be 0.93 mm from the minimum 2-norm value. Figure 1(c) shows the projection data in positive (solid line) and negative directions (dotted line) after deconvolution using (10) with $\xi_c$ of 0.93 mm, showing that the two sets of projection data almost overlap.

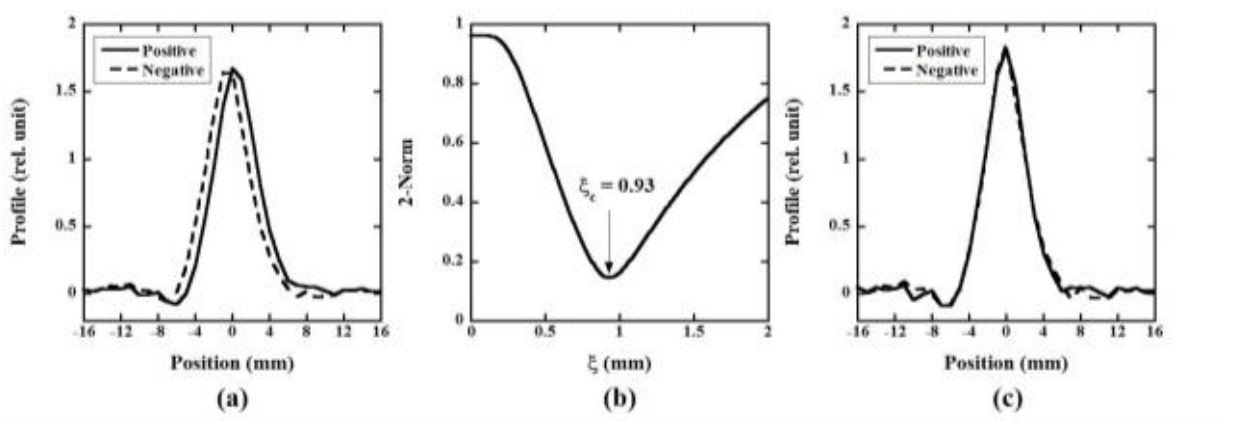

**Figure 1:** (a) Projection data acquired in positive (solid line) and negative directions (dotted line) for $v_{FFL}$ = 2.0 mm/s. (b) 2-norm values as a function of $\xi$. (c) Projection data in positive (solid line) and negative directions (dotted line) after deconvolution using $\xi_c$ of 0.93 mm.

Figure 2 shows the correlation between $v_{FFL}$ and $\xi_c$. There was a significant linear correlation between them.

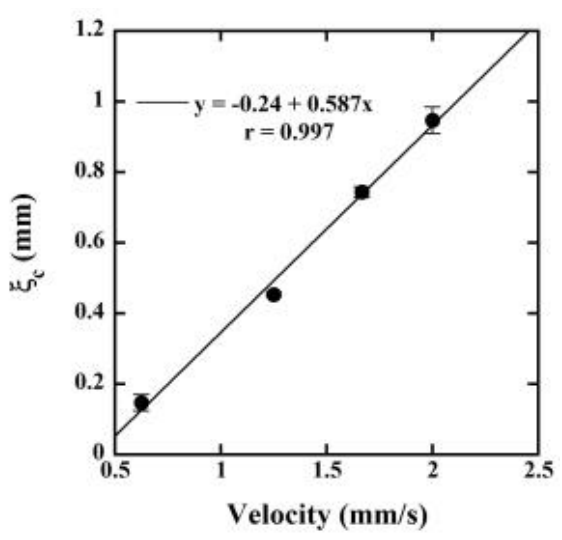

**Figure 2:** Correlation between $v_{FFL}$ and the $\xi_c$ value. Error bar represents standard deviation (SD) for n=3.

Figure 3 shows an example of the sinograms of a line phantom before (upper row) and after correction of the signal delay (lower row) for various values of $v_{FFL}$ ((a) for 0.62 mm/s, (b) for 1.25 mm/s, (c) for 1.67 mm/s (c), and (d) for 2.0 mm/s). The shift of projection data due to a difference in the scanning direction of FFL increased with increasing $v_{FFL}$, while these shifts disappeared after correction of the signal delay.

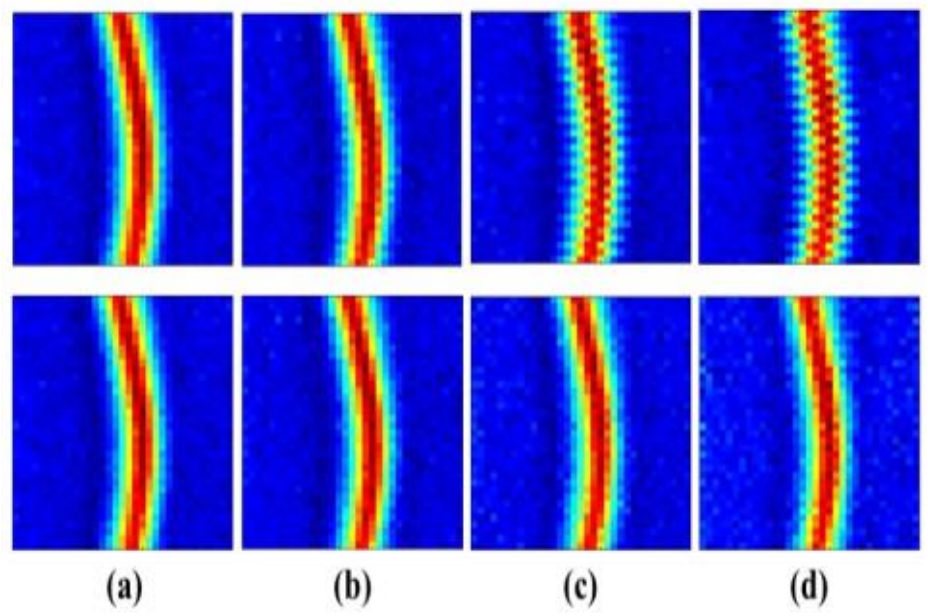

**Figure 3:** Signograms of a line phantom before (upper row) and after correction of the blur (lower row) for $v_{FFL}$ = 0.62 mm/s (a), 1.25 mm/s (b), 1.67 mm/s (c), and 2.0 mm/s (d). The vertical and horizontal axes in the sinogram represent each projection angle and the distance along the projection direction, respectively.

Figure 4(a) shows an example of the MPI image of the line phantom for $v_{FFL}$ = 2.0 mm/s before correction of the blur, while Fig. 4(b) shows that after correction of the blur using 0.95 mm as $\xi_c$. Figure 4(c) shows the horizontal profiles of the MPI image of the line phantom before (solid line) and after correction of the blur (dotted line), while Fig. 4(d) shows the case of the vertical profile. Note that the lines along which the horizontal and vertical profiles were obtained are shown by solid lines in Figs. 4(a) and 4(b).

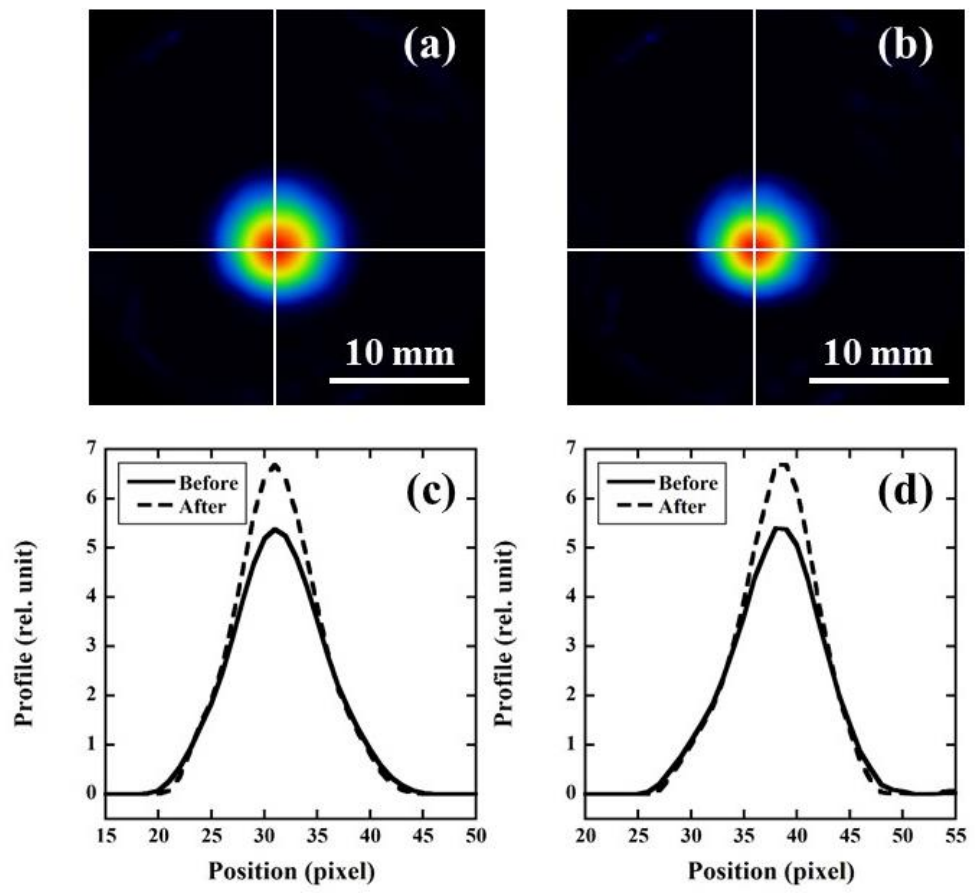

**Figure 4:** MPI images of a line phantom before (a) and after correction of the blur (b) for $v_{FFL}$ = 2.0 mm/s. Horizontal (c) and vertical profiles (d) before (solid line) and after correction of the blur (dotted line). Scale bar = 10 mm.

Figure 5(a) shows the horizontal FWHM values before (closed circles) and after correction of the blur (open circles) as a function of $v_{FFL}$, while Fig. 5(b) shows the case of the vertical FWHM. Both the horizontal and vertical FWHM values increased with $v_{FFL}$. There were significant differences in both the horizontal and vertical FWHM values between before and after correction of the blur for $v_{FFL}$ = 1.67 mm/s and 2.0 mm/s.

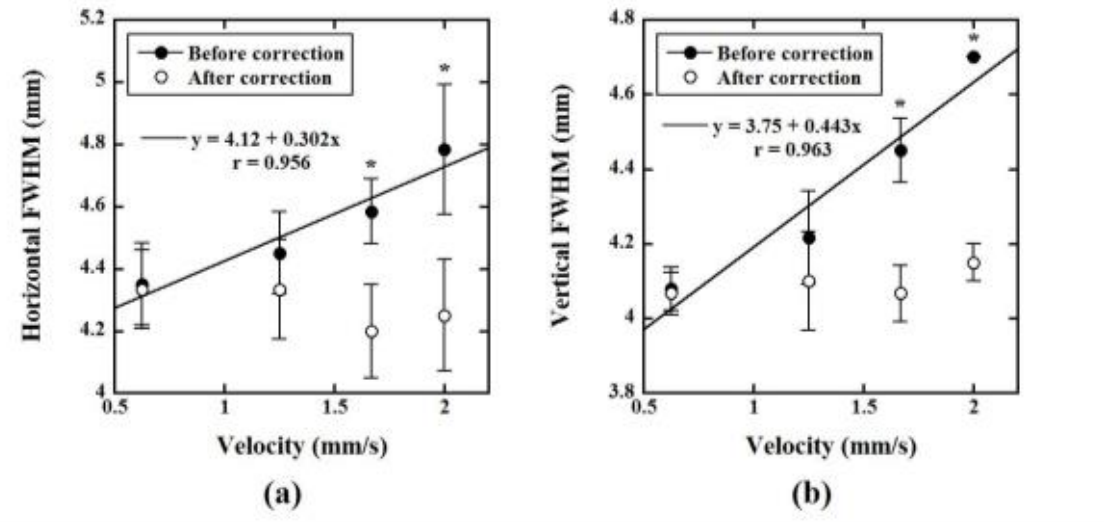

**Figure 5:** Horizontal (a) and vertical FWHM values (b) as a function of $v_{FFL}$ before (closed circles) and after correction of the blur (open circles). Error bar represents SD for n=3. * $P<0.05$.

Figure 6 shows a comparison between the MPI images of the A-shaped phantom before (upper row) and after correction of the blur (lower row) for various values of $v_{FFL}$ ((a) for 0.62 mm/s, (b) for 1.25 mm/s, (c) for 1.67 mm/s (c), and (d) for 2.0 mm/s). The blur in the MPI image increased with increasing $v_{FFL}$ (upper row), while the effect of the correction was more clearly observed with increasing $v_{FFL}$ (lower row).

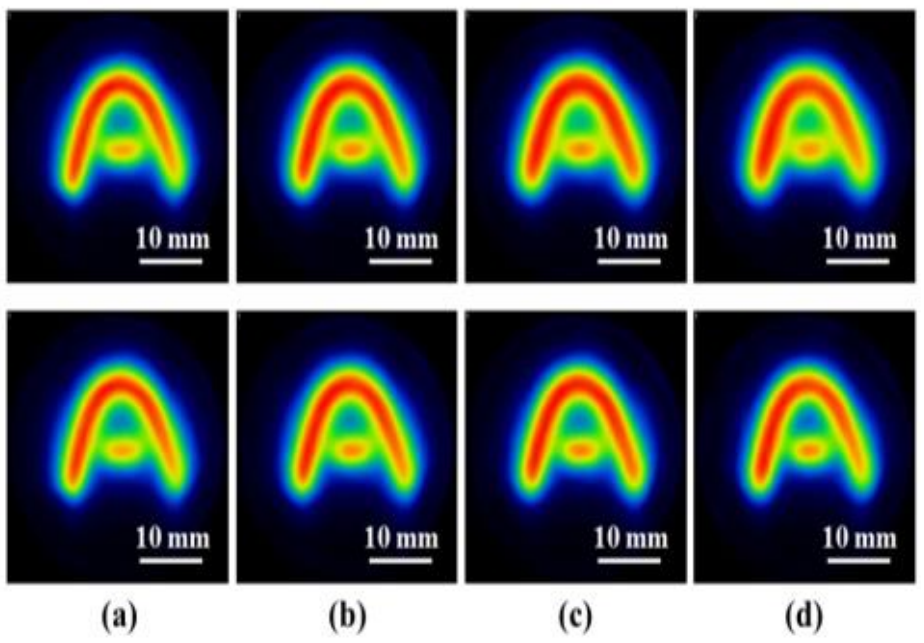

**Figure 6:** MPI images of an A-shaped phantom before (upper row) and after correction of the blur (lower row) for $v_{FFL}$ = 0.62 mm/s (a), 1.25 mm/s (b), 1.67 mm/s (c), and 2.0 mm/s (d). Scale bar = 10 mm.

## 4. Discussion

When we consider the practical application of projection-based MPI with an FFL-encoding scheme, it is necessary to acquire projection data by moving the FFL as fast as possible, and thus projection data are usually acquired in a zigzag. This may cause a difference in the projection data depending on the scanning direction of the FFL. We then investigated whether a shift of projection data occurs depending on the scanning direction of the FFL using our MPI scanner. Our results of phantom experiments supported the existence of such a shift in the projection data, resulting in the blur in the reconstructed images. In addition, we developed a method for correcting the blur due to the above shift by deconvolution based on SVD with a correction factor determined from the projection data acquired in positive and negative directions. Our results demonstrated the validity and usefulness of our method.

As shown in Fig. 2, there was a linear correlation between $v_{FFL}$ and $\xi_c$, suggesting that $\tau$ in (1) is constant regardless of $v_{FFL}$, because $\xi$ is given by the product of $\tau$ and $v_{FFL}$. From the slope of the regression line in Fig. 2, $\tau$ was estimated to be approximately 500 ms. This value is much larger than the magnetic relaxation time of MNPs [4]. Thus, the signal delay may be mainly due to other reasons such as the delay in the data acquisition including analog-to-digital conversion and the control system. Further studies will be necessary to elucidate the meaning of this parameter.

As shown in the lower row of Fig. 3, the noise in the sinogram after deconvolution increased with increasing $\xi_c$ value. As previously described, the threshold value used in SVD was fixed at 0.1 in this study. When a larger value of this parameter was used, the noise level was more suppressed (data not shown). Thus, it would be possible to control the noise level in the sinogram after deconvolution and the resulting reconstructed MPI image by adjusting this parameter according to the noise level in the sinogram [5].

In this study, $v_{FFL}$ was varied as 0.62, 1.25, 1.67, and 2.0 mm/s. Although we could not study cases with $v_{FFL}$ greater than 2.0 mm/s because of the limitations of our hardware, we believe that our method will also be applicable to projection-based MPI with greater $v_{FFL}$.

Although our method requires additional time to perform two scans in positive and negative directions for obtaining the $\xi_c$ value, this step is not always necessary. Thus, the time required to perform the two scans may not be a significant hindrance in practical use. In addition, our method will be advantageous in terms of cost, because it does not require any additional hardware or the modification of existing hardware.

In conclusion, we developed a method for correcting the blur caused by a difference in the scanning direction of FFL. Our results suggest that our method will be useful for improving and enhancing the reliability of projection-based MPI.